\documentclass[sigplan,10pt]{acmart}
\renewcommand\footnotetextcopyrightpermission[1]{}
\usepackage{multirow}

\usepackage{listings}
\usepackage{graphics}
\newcommand{\verbatimfont}[1]{\renewcommand{\verbatim@font}{\ttfamily#1}}
\usepackage{subcaption}
\usepackage{graphicx}

\newcommand{\ignore}[1]{}
\newcommand{\tname}{CGuard}
\newcommand{\toolname}{CGuard }
\newcommand{\piyus}[1]{#1}

\author{Piyus Kedia} 
\affiliation{%
  \institution{IIIT Delhi}
  \country{India}
}

\author{Rahul Purandare} 
\affiliation{%
  \institution{UNL}
  \country{USA}
}

\author{Udit Agarwal} 
\authornote{A part of this work was done while the author was at IIIT Delhi}
\affiliation{%
  \institution{UBC}
  \country{Canada}
}

\author{Rishabh}
\affiliation{%
  \institution{GGSIPU}
  \country{India}
}

\begin{abstract}
Spatial safety violations are the root cause of many security attacks and unexpected behavior of applications. Existing techniques to enforce spatial safety work broadly at either object or pointer granularity. Object-based approaches tend to incur high CPU overheads, whereas pointer-based approaches incur both high CPU and memory overheads.
SGXBounds, an object-based approach, provides precise out-of-bounds protection for objects at a lower overhead compared to other tools with similar precision. However, a major drawback of this approach is that it cannot support address space larger than 32-bit.

In this paper, we present \tname, a tool that provides precise object-bounds protection for C applications with comparable overheads to SGXBounds without restricting the application address space. \toolname stores the bounds information just before the base address of an object and encodes the relative offset of the base address in the spare bits of the virtual address available in x86\_64 architecture. For an object that cannot fit in the spare bits, \toolname uses a custom memory layout that enables it to find the base address of the object in just one memory access. Our study revealed spatial safety violations in the {\tt gcc} and {\tt x264} benchmarks from the SPEC CPU2017 benchmark suite and the {\tt string\_match} benchmark from the Phoenix benchmark suite. The execution time overheads for the SPEC CPU2017 and Phoenix benchmark suites were 42\% and 26\% respectively, whereas the reduction in the throughput for the Apache webserver when the CPUs were fully saturated was 30\%. These results indicate that \toolname can be highly effective while maintaining a reasonable degree of efficiency.
\end{abstract}

\begin{document}

\date{}

\title{\Large \bf \tname: Efficient Spatial Safety for C}

\settopmatter{printfolios=true}
\settopmatter{printacmref=false}
\maketitle

\pagestyle{plain}

\section{Introduction}
Spatial safety violations are the root cause of many security attacks \cite{ropsecurity,jopsecurity,smashing-security,returnlibcsecurity,rop1security,missingsecurity,breakingsecurity,losingsecurity,guardsecurity}. Attackers can exploit spatial safety bugs to hijack an application's control flow or steal sensitive information (e.g., passwords). Beyond security issues, spatial safety is 
important 
to ensure expected application behavior. 
For example, unintentionally accessing an out-of-bounds location can cause unexpected behavior or program crashes that are hard to debug. 

Spatial safety is just one aspect of reliability. Managed languages, such as Java and C\#, offer better reliability by providing complete (spatial and temporal) memory and type safety. However, C does not guarantee any of these safeties by default. Despite the lack of memory safety, C and C++ are still preferred over managed languages for systems applications because managed languages are less efficient. Consequently, many performance-sensitive applications are still vulnerable to security exploits and are therefore not reliable. In this work, we propose a mechanism to enforce spatial safety for C applications.

Several techniques have been proposed to enforce spatial safety for C/C++ applications. At a high level, these techniques can be categorized into pointer-based \cite{ccured,condit2003ccuredreal,softbound,cyclone,SafeC,xu2004efficient} and object-based \cite{joneskelly,CRED, CRED++, sgxbounds, Baggy, ParicheckAlloc, LFPCC, LFPNDSS} approaches.

Pointer-based approaches track the bounds of sub-objects and can detect sub-object overflows. 
Even with hardware support \cite{hardbound,MPX}, these approaches incur high CPU and memory overheads because they need to store and update bounds information for every pointer. Oleksenko et al. \cite{oleksenko2018intel} have reported around 75\% CPU and 125\% memory overheads for SPEC benchmarks for the Intel MPX \cite{MPX} implementation of the ICC compiler. 

In object-based approaches, spatial safety checks ensure that the memory access using a pointer is within the heap/stack/global allocation bounds. These approaches have low memory overheads because they do not need to store the bounds for every pointer. Finding the base or limit of an object using a pointer is challenging because the pointer can be an internal  address of an object. Initial approaches \cite{joneskelly,CRED,CRED++} used a splay-tree-based lookup to check if the pointer points to a location within object bounds. For efficiency, later works \cite{Baggy,ParicheckAlloc,LFPCC,LFPNDSS} enforced spatial safety at loose (or imprecise) allocation bounds rather than the actual allocation bounds. These works pad the actual allocation size to satisfy an alignment property and keep track of alignment instead of the actual allocation size. 
An important drawback of these approaches is that they allow applications to access the padded area that is not within the actual bounds of the objects. This would allow unintended behavior that may remain undetected.

Our goal is to enforce checking for the actual bounds of the object, thereby providing \textit{precise} object-bounds protection.
SGXBounds \cite{sgxbounds} is so far the most efficient technique (41\% and 55\% CPU overheads inside and outside the SGX enclaves \cite{SGX,mckeen2013innovativesgx} for SPEC CPU2006) that provides precise object-bounds protection, but it restricts the application's usable address space to 32-bit on a 64-bit platform. SGXBounds uses the remaining 32 bits to store the upper bound of the object. This allows SGXBounds to compute the upper bound directly from the pointer itself without any expensive search. The fundamental weakness of this approach is that it cannot support larger address space because the upper bound cannot fit in the unused bits of the virtual address. As a consequence, the application's address space gets restricted to 32-bit.

\begin{figure}
\centering
\includegraphics[width=0.44\textwidth,height=1.1cm]{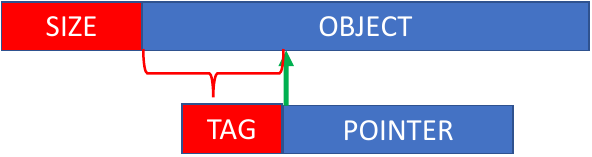}
\caption{Object layout (top) and pointer layout (bottom) in a general case. Here, the pointer is pointing to some internal field in the object (shown by the arrow). The tag contains the relative address with respect to the base address of the object. The size is stored just before the base address.}
\label{fig:layout}
\end{figure}

We propose \tname, a tool that provides object-bounds protection without restricting the virtual address space.
Figure~\ref{fig:layout} shows the layout of an object and pointer in our scheme in a general case. \toolname stores the size of an object before the base address of the object and attaches a tag to every pointer to efficiently locate the base address. \toolname uses the spare 16-bits of a virtual address available in the {\tt x86\_64} hardware to store the tag. In the tag, \toolname stores the relative offset of the pointer with respect to the base address of the object referred to by the pointer. To find the base address, \toolname simply subtracts the offset from the pointer value. For objects that cannot fit in the spare bits, \toolname uses a custom allocator that allows it to find the base of an object in just one memory access. A major challenge in our design is that, unlike SGXBounds, \toolname needs to update the offset in the tag on every pointer arithmetic. \toolname performs static analysis to reduce the number of tag updates. The mean CPU and memory overheads incurred by \toolname for SPEC CPU2017 are 42.1\%  and 1.1\%, respectively.

Spatial safety mechanisms for managed languages are well understood. The size of an object is stored along with the object. Managed languages do not allow pointer arithmetic, enabling the mechanisms to discover the size of the object at all program points statically. On the other hand, C allows programmers to create interior pointers, store them in memory, pass them to other routines, and return them to a caller. This makes the static tracking of base pointers very hard. In our approach, the tag information needs to be updated only if a statically known potential interior pointer escapes the static scope. Thus our scheme allows programmers to control the overhead of spatial safety.
If the usage of interior pointers is restricted to the static scope, our tagged pointers are equivalent to normal pointers, and the spatial safety handling mechanism is similar as in the case of a managed language.

\piyus{AddressSanitizer \cite{addresssanitizer} detects sequential overflows and underflows; and some \textit{use-after-free} bugs at low overhead. The tool has already been integrated into GCC and LLVM compilers. It tracks the validity of stack, heap, and global objects using {\tt shadow memory}. For every eight-byte of main memory, one byte of shadow memory is used to track its validity. The shadow memory is kept at a constant offset from the corresponding main memory, and thus checking the validity of a memory address before the access is very efficient. To detect overflows, AddressSantizer inserts extra memory blocks, {\it redzones}, around every object, marks them invalid in the shadow memory, and then tracks access to them. However, this scheme cannot detect out-of-bounds accesses that jump the redzones. 
AddressSanitizer detects use-after-bugs by putting the freed region into {\tt quarantine} for some time. If access to a free object happens during quarantine, then it's a use-after-free bug. We compared \toolname also with AddressSanitizer because it performs better than SGXBounds in a normal unconstrained environment.}

In summary, we make the following contributions.
\begin{enumerate}
    \item An approach based on pointer tagging to provide object-bounds protection for C applications at low overheads without restricting the application address space.
    \item An optimization and its evaluation that eliminates the need for bounds checking for structure accesses that are used similar to objects in managed languages. 
    \item A tool, \tname, based on LLVM and its performance evaluation using real world benchmarks.
    \item Detection and reporting of bugs in the SPEC CPU2017 and Phoenix-2.0 benchmark suites.
\end{enumerate}


\section{Design}
\label{sec:design}

\subsection{Overview}

\begin{figure}
\begin{lstlisting}
1.  int* bar(int *arr, int i, int **var,
2.           struct node *n) {
3.    int *newarr = *var;
4.    arr[i] = 200;
5.    newarr[i] = 40;
6.    if (newarr == arr + 1)
7.      n->field_i = 0;
8.    return &arr[i];
9.  }
10. void foo(int i, int **var) {
11.   int x[100];
12.   struct node n;
13.   *var = &x[6];
14.   *var = bar(&x[5], i, var, &n);
15. }
\end{lstlisting}

\caption{Code snippet to discuss the overview of \tname.}
\label{fig:overview}
\end{figure}


\toolname inserts an {\it object-header} before the object's base address to store the object's size. To compute the object bounds, we need to find the base address of the object at runtime. 

In the code snippet depicted in Figure~\ref{fig:overview},
the argument {\tt arr} in {\tt bar} is an interior pointer. To compute the bounds of {\tt arr} at line-4, we need to find the base address of {\tt arr}. To locate the base address, we store the offset from the base address in the tag area of the pointer. Here, the tag area of argument {\tt arr} contains value 20. Using this information, \toolname can compute the base address by simply subtracting offset (20) from the virtual address of {\tt arr}. \toolname does not update the tag area for every pointer. For example, at line-4, after computing the address of {\tt arr[i]} for the memory access, \toolname does not need to update the tag because it statically knows that {\tt arr} and {\tt \&arr[i]} belong to the same array, and it can compute the base address from the argument {\tt arr} at line-1. We call {\tt arr} the {\it static-base} of {\tt \&arr[i]}. Similarly, the static-base of {\tt \&newarr[i]} at line-5 is {\tt newarr} at line-3.
\toolname statically analyzes the routine to identify the static-base for every pointer.
Whenever a pointer escapes the static scope, it may become a static-base in other parts of the program. For example, at line-13, {\tt \&x[6]} leaves the static scope and becomes the static-base at line-3. Therefore, we update the tag before storing {\tt \&x[6]} in {\tt var} at line-13. Similarly, \toolname updates the tag of {\tt \&x[5]} (at line-14) and {\tt \&arr[i]} (at line-8) before passing to and returning from the {\tt bar} routine. \toolname does not need to update the tag while storing the return value of {\tt bar} in {\tt var} at line-14. This is because the return value of a function is a static-base, and it already has the correct offset in its tag area.

A problem with this approach is that the maximum offset gets restricted by the number of bits in the tag field (\toolname uses 15 bits to store the offset). For objects that cannot fit into 15 bits, \toolname uses a segmented heap. In this case, the base of the object is computed using the alignment property of the segmented heap. Another problem is that C does not distinguish between a pointer and an array. For example, argument {\tt n} in {\tt bar} at line-2 is a pointer to a structure element; however, \toolname needs to add bounds check at line-7 before accessing the structure field because it could be an array of structures. On the contrary, in managed languages, the type-system can distinguish between an object and an array of objects. Therefore, object accesses do not need to perform explicit bounds checks. To eliminate the need for these bound checks, 
we expect all static-bases to point to a memory area that is large enough to store at least one element of the corresponding array. We call this property the {\it size-invariant} property. 
In the above example, \toolname requires the argument {\tt n} to point to a memory area that is at least ``{\tt sizeof(struct node)}'' long. We found that for most benchmarks, this property 
holds. In our scheme, programs that do not satisfy 
this
property may have to pay an additional performance penalty. 

In our scheme, changing the pointer layout further complicates the pointer comparison and subtraction operations. Now, the same pointers may have different offsets in their tag areas depending on their static bases. For example, the equality checks at line-6 will fail because {\tt newarr} and {\tt arr} contain different offsets in their tag areas. To handle this correctly, \toolname resets the tags in the pointer operands during these operations. \toolname also resets the tag before every memory access. \toolname uses custom wrappers to invoke system library routines. These wrappers reset the tag field from the pointer arguments because the unmodified library does not understand \tname's pointer layout. Finally, \toolname inserts dynamic checks before memory accesses to abort the program if the accesses are not within the object-bounds.

\begin{figure}
\centering
\includegraphics[width=0.44\textwidth,height=3.4cm]{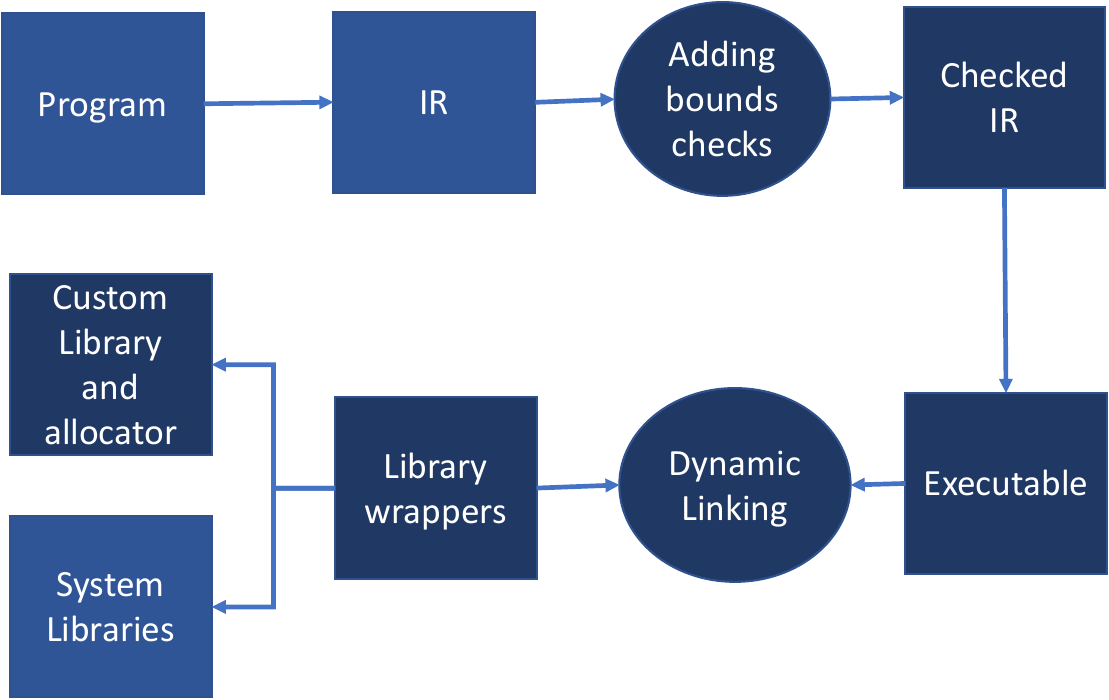}
\caption{Architecture of \tname.}
\label{fig:architecture}
\end{figure}

Figure~\ref{fig:architecture} shows the architecture of \tname. \toolname takes the intermediate representation (IR) of a program as input. The IR is in static single assignment (SSA) form. We incorporate our spatial safety logic in the IR to generate the checked IR. The checked IR is compiled to an executable. At load time, the executable is linked with our custom library that implements wrappers, custom library routines, and the custom allocator.
In the rest of this section, we explain our scheme in detail.

\subsection{Identifying Static-Base}
\label{sec:static_base}
Before every dereference of pointer x, the bounds can be computed using the offset field in the tag area of x. Ensuring the correct offset 
for every pointer definition is expensive because tags need to be updated on every pointer arithmetic. To reduce the number of 
updates, when pointer x is dereferenced, \toolname statically tries to find another pointer 
y that points to the same object as x and contains the correct offset. 
We call y the static base of x. The bounds of the object are computed using y.
We explain below our approach 
to find the static-base for different kinds of definitions in the IR. \ignore{\rahul{I think it would be better to give an algorithm at least in the form of pseudocode here.} \piyus{I have added an example for typecast and the pointer arithmetic instructions. For ptrtoint it is hard to give a pseudocode.} \rahul{Okay. Will take a look. However, in that case, I suggest we should avoid using the word ``algorithm''. People will expect either a formal algorithm or at least some pseudocode in that case. We can just say that ``we explain below our approach to find the static-base''. In general, I would say, providing algorithm or pseudocode for such important operations will help. In addition to changing words, we should also enumarate cases here. That will provide more clarity about the approach.}.
\piyus{Okay, we can discuss this in our next meeting. The problem is the pointer-to-integer operations are not sound, and therefore it is hard to write a clear algorithm. Also, I think the main problem with the static-base is that the people do not understand the need of it. I have added a discussion in the first paragraph that would perhaps make it clear.}
}
\\
1) For pointer arithmetic and typecast operation {\tt x}, 
we recursively trace back all arithmetic and typecast operations to obtain a pointer {\tt y} that is not the result of pointer arithmetic or a typecast operation. In this case, the static-base of {\tt y} is the static-base of {\tt x}. Consider the following IR code.

\begin{verbatim}
x = bitcast ty1 ptr to ty2
y = getelementptr ty, ty* arr, i32 i
\end{verbatim}

\ignore{The above code shows instructions corresponding to typecast and pointer arithmetic operations in an IR.} Here, the {\tt bitcast} instruction generates a new definition {\tt x} after casting {\tt ptr} of type {\tt ty1} to {\tt ty2}. In this case, the static-base of {\tt x} is the same as the static-base of {\tt ptr}. The operands of {\tt getelementptr} instruction are {\tt arr} and {\tt i}. {\tt getelementptr} generates a new definition {\tt y} whose value is {\tt \&arr[i]}. In this case, the static-base of {\tt y} is the same as the static-base of {\tt arr}.
\\
2) For an integer-to-pointer typecast {\tt x}, if we can statically correlate {\tt x} with a previous pointer-to-integer operation {\tt y}, we infer the static-base of {\tt x} as the static-base of {\tt y}. If 
such a
pointer-to-integer operation is not found, {\tt x} is treated as the static-base of itself.
\\
3) Pointers loaded from memory, the return value of a function call, function arguments, stack allocations, and global variables are also the static-bases of themselves.
\\
4) The SSA representation contains {\it phi-nodes} to merge the definitions coming from multiple predecessor basic-blocks. In this case, we add a new phi-node that merges the static-bases of the definitions coming from these predecessors.

\begin{verbatim}
z_sb = phi <sb(x), pred1>, <sb(y), pred2>
z = phi <x, pred1>, <y, pred2>
\end{verbatim}

In this example, {\tt z} is a phi-node that merges definitions {\tt x} and {\tt y} coming from basic blocks {\tt pred1} and {\tt pred2}. 
We add a new phi-node {\tt z\_sb}, the static-base of {\tt z}, that merges the static-bases of {\tt x} and {\tt y} denoted using {\tt sb(x)} and {\tt sb(y)}. 
\\
5) The IR contains the instruction {\tt select} that emulates the ternary operator as shown below.

\begin{verbatim}
z_sb = select cond, sb(x), sb(y)
z = select cond, x, y
\end{verbatim}

In this example, {\tt select} takes condition {\tt cond} and definitions {\tt x} and {\tt y} as input and creates a new definition {\tt z}. At runtime, {\tt z} will be equal to {\tt x} or {\tt y} depending on the value of {\tt cond}. To find the static-base, we add an additional {\tt select} instruction that takes {\tt cond}, {\tt sb(x)}, and {\tt sb(y)} as inputs and create a new definition {\tt z\_sb}, the static-base of {\tt z}. 

\subsection{Tagged Pointer}
\label{sec:tag_ptr}

We use the spare higher 16-bits of virtual address 
to store the tag. 
A tagged-pointer has the following structure.

\begin{lstlisting}
typedef struct {
  unsigned long long address:48;
  unsigned long long invalid:1;
  unsigned long long offset:15;
} tag_t;
\end{lstlisting}

In the rest of the paper, we will refer to the tagged-pointer type using {\tt tag\_t}. The lower 48-bits of a pointer contain the actual address, represented using the address field in the {\tt tag\_t}. The invalid field is used to mark a pointer invalid (as discussed later in this section).
The maximum offset that can be stored in the 15-bits offset field is MAX\-\_OFFSET (0x7FFF). The allocation for a size larger than or equal to MAX\_OFFSET is performed from the segment-based allocator (discussed in the next paragraph).
The offset field in the static-base tagged pointer contains the offset relative to the actual base address of the object referred by the static-base.
If the relative offset is equal to MAX\_OFFSET, then the base address is computed using the alignment property of the segment-base allocator.

\ignore{
\label{sec:alloctor}
\begin{figure}[!t]
\centering
\includegraphics[width=3cm, height=3cm]{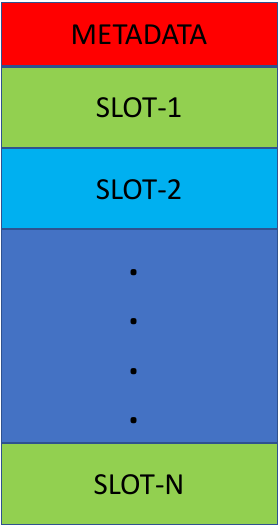}
\caption{Layout of a segment.}
\label{fig:segment}
\end{figure}
}


\paragraph{{\bf Segment-based allocator}} The segment-based allocator maintains a list of segments that are shared across all threads. A segment 
is a 4GB (configurable at compile time) contiguous virtual address space. The starting address of a segment is aligned to 4GB. The segment is divided into fixed-size slots. Both the size and alignment of a slot are $2^k$ (a power of two). The value of {\tt k} can vary across segments.
The first few pages of the segment are used to store the metadata (e.g., a bitmap to track free slots). Initially, the virtual addresses are reserved for the entire segment. Physical pages are mapped only during the actual allocation. For every allocation,  a slot is returned to the caller. Because a slot can be much larger than an actual allocation size, we only map the number of physical pages that are sufficient for the allocation size. The physical pages are reclaimed during the deallocation.

\toolname manages stack allocations of sizes greater than or equal to MAX\_OFFSET using {\tt malloc} and {\tt free}. For these objects, \toolname replaces the calls to stack allocation API with calls to {\tt malloc} and inserts {\tt free} when the objects go out of scope. \toolname also inserts object-headers before stack and global allocations.

\paragraph{{\bf Updating the pointer tag}}
The tag is updated every time a pointer escapes the static scope 
as a result of it being passed to a function, stored in memory, or returned to a caller. We do not track a pointer if it escapes after being typecasted to an integer. Instead, we expect that the program casts it back to a pointer before the escape if the integer is out-of-bounds or modified due to some arithmetic operations on the integer. After the escape, the pointer may become a static-base in other parts of the program.  For example, in our static-base identification logic (Section~\ref{sec:static_base}), a loaded value is identified as static-base. After a pointer is stored in memory, it can be loaded at different parts of the program and treated as a static-base. We update the 
tag before the escape to ensure that all the tagged static-bases always contain the correct offset. If the pointer is not within the bounds or does not satisfy the size-invariant property (Section~\ref{sec:size_inv}), the invalid bit in the tag area is set. Accessing pointers with the invalid-bit set result in runtime exceptions preventing out-of-bounds memory accesses. 

\ignore{The {\tt update\_tag} routine in Figure~\ref{fig:update_metadata} updates the tags for escaping pointers. In this routine, {\tt base} is the actual base address of the object, {\tt ptr} is the escaping pointer, {\tt access\-\_size} is the size of the type of array element that {\tt ptr} is pointing to, and the {\tt limit} is the upper bound of the object. The return value of {\tt update\_tag} is a tagged pointer (for the address in {\tt ptr}) with the correct offset.
The {\tt update\_tag} routine stores the relative offset of the {\tt ptr} address with respect to the {\tt base} address in the return value if the relative offset is less than or equal to {\tt MAX\-\_OFFSET}. Otherwise, {\tt MAX\_OFFSET} is stored in the return value. If the address of {\tt ptr} is not within the bounds, the invalid bit is set in the return value. Notice that {\tt update\_tag} is only needed if {\tt ptr} is not a statically known alias of the static-base of {\tt ptr}, or if {\tt access\_size} is larger than the size of the array element type pointed by the static-base of {\tt ptr}. The {\tt access\_size} check is needed for the {\it size-invariant} property discussed in Section~\ref{sec:size_inv}. \\
}
\paragraph{{\bf Handling out-of-bounds pointers}}
In the general case, \toolname allows memory access when an in-bounds pointer {\tt y} derived from an out-of-bounds static-base {\tt x} (due to pointer arithmetic) is dereferenced. If the offset field in the tag area of {\tt x} is less than {\tt MAX\_OFFSET}, the actual base address can be computed by subtracting the offset and ignoring the invalid-bit. In the other case, \toolname cannot compute the actual base, and thus the dereference of {\tt y} is not allowed.

\subsection{Computing Bounds and Inserting Checks}
\label{sec:base_comp}
\begin{figure}
\begin{lstlisting}
#define OBJ_HEADER_SIZE 8
#define SEGMENT_MASK ~(SEGMENT_SIZE - 1)

void *get_base(tag_t sb) {
  if (sb.offset < MAX_OFFSET)
    return (void*)(sb.address - sb.offset);
  if (sb.invalid) return NULL;
  if (is_global_var(sb.address))
    return get_base_allocator(sb.address);
  segment_t *s;
  void *ret;
  s = (segment_t*)(sb.address & SEGMENT_MASK);
  ret = (void*)(ptr.address & s->slot_mask);
  return ret + OBJ_HEADER_SIZE; 
}
\end{lstlisting}
\caption{\label{fig:get_base} Routine used to obtain the base from an input tagged static-base ({\tt sb}). If the offset in the tagged static-base is less than MAX\_OFFSET, the base (lower bound) is computed after subtracting the offset from the address of the static-base pointer. Otherwise, the segment alignment property or the allocator API is used to obtain the base.}
\end{figure}

\toolname computes the base address of a pointer definition using the tagged static-base. The base computation logic ({\tt get\_base}) is shown in Figure~\ref{fig:get_base}. If the offset is less than {\tt MAX\_OFFSET}, {\tt get\_base} subtracts the offset in the tag from the address of the static-base pointer. Otherwise, if the static-base is invalid (i.e., out-of-bounds), {\tt get\_base} cannot retrieve the actual base and returns {\tt NULL}. If the offset is equal to the {\tt MAX\_OFFSET} and the address belongs to the range of global variables, it calls the allocator API {\tt get\_base\_allocator} (discussed in Section~\ref{sec:implementation}), 
which
does not rely on pointer tag to obtain the base. It also maintains a small cache to avoid calls to the allocator API, which works well in practice because there are only a few global variables of size greater than or equal to MAX\_OFFSET across all of our benchmarks.
Finally, for segment-based allocation, the base address of the object is computed using the alignment property of the segments. 
All slots in a segment are aligned to $2^k$. The starting address of a slot 
is computed by resetting the lower k-bits of the pointer address. The first eight bytes of a segment contain {\tt slot\_mask} (${\sim}(2^k-1)$). The starting address of the object slot is computed by `anding' the pointer address and the {\tt slot\_mask}. The starting address of a slot is the object-header. The base address is computed by skipping the 
header.

If the static-base is an integer-to-pointer typecast, the offset field can be incorrect due to untracked integer arithmetic operations performed on the integer. To handle this case, if the static-base is an integer-to-pointer instruction or a {\tt phi} or {\tt select} node that depends on an integer-to-pointer instruction, \toolname backtracks all operations on the integer to check if it is involved in any arithmetic. If so, \toolname uses the allocator API to find the base.
In case an integer which escapes the static scope with an incorrect tag can be accessed in the future, we rely on the application to typecast it into a pointer before letting it escape.

\paragraph{{\bf Bounds check}}
Our bounds check logic is shown below. 
\begin{lstlisting}
void bounds_check(void *base, void *ptr,
    void *ptrlimit, void *limit) {
  if (ptr < base || ptrlimit > limit) abort();
}
\end{lstlisting}
The arguments to the {\tt bounds\-\_check} routine are the pointer ({\tt ptr}) (without tag) that is being accessed, the base address ({\tt base}) of the object referred by {\tt ptr} (e.g., obtained using {\tt get\-\_base}), the upper bound of the memory access ({\tt ptrlimit}), and the upper bound of the object ({\tt limit}). The upper bound of the object is computed by adding 
its size obtained from the object-header
to its base address. 
If {\tt ptr} does not lie between {\tt base} and {\tt limit}, the program is aborted. 

\subsection{Size-Invariant}
\label{sec:size_inv}
\toolname enforces the size-invariant to eliminate checks when only the first element of an array or pointer to a structure element is accessed. This invariant requires all static-bases to point to a memory area that is large enough to store at least one element of the corresponding array. For example, if {\tt char *a} is a static base, then {\tt a} must point to a memory area that is at least one byte long; if the type of {\tt a} is  {\tt unsigned long long *}, it must point to a memory area that is at least eight bytes long. If a pointer escapes the static scope, we invalidate the pointer if the size-invariant does not hold.
\ignore{The corresponding logic is 
depicted in
(Figure~\ref{fig:update_metadata}). Here, {\tt access\_size} is the size of the array element type. If {\tt ptr.address} and {\tt ptr.address + access\_size - 1} are not within the object-bounds, the invalid bit in the tag is set.} This allows \toolname to remove bounds-check when only the first element of the array is accessed, since \toolname does not reset the invalid bit for these accesses. Accessing a pointer that does not satisfy the size-invariant property result into access violation.

In our experiments, we found that the size-invariant holds for the majority of the benchmarks (Section~\ref{sec:usability}). For the benchmarks that violate the invariant, the problem can be addressed either by using a smaller type for the static-base and external typecasts whenever needed or by 
extra allocation. Consider the following code snippet:
\begin{lstlisting}
struct info {
  int a, b, c, d;
};
int foo(struct info *i) {
  return i->b;
}
void bar() {
  int arr[2] = {1, 2};
  return foo((struct info*)arr);
}
\end{lstlisting}

In this snippet, the size-invariant requires {\tt bar} to pass an object of size at least {\tt sizeof(struct info)} to {\tt foo}. Because the size-invariant does not hold, \toolname invalidates the parameter passed to {\tt foo}. The hardware generates an access violation when {\tt foo} tries to dereference the invalid pointer. In this case, the bounds check 
is
performed in the signal handler as discussed in 
Section~\ref{subsec:recovery}. However, signal handling is expensive. These cases can be efficiently handled using code refactoring.
A way to fix this problem is to allocate at least {\tt sizeof(struct info)} memory for the variable {\tt arr} in the {\tt bar} routine. This approach incurs memory overhead. 
An alternative is to rewrite the {\tt foo} and {\tt bar} routines as follows:

\begin{lstlisting}
int foo(int *a) {
  struct info *i = (struct info*)a;
  return i->b;
}
void bar() {
  int arr[2] = {1, 2};
  return foo(arr);
}
\end{lstlisting}

In this case, since the argument type in {\tt foo} is {\tt int*}, {\tt bar} does not invalidate the parameter passed to {\tt foo}.
\toolname adds dynamic checks in {\tt foo} while dereferencing {\tt i} because based on 
the size-invariant, it only knows that {\tt i} is at least four bytes long.
This approach does not incur any memory overhead but has a CPU overhead due to bounds checking. 
If types cannot be modified, an additional 
attribute can be used to disable or pick a different size for the size-invariant optimization for a given type. We plan to implement the type attribute in the future.

\subsection{Recovery from Size-Invariant Errors}
\label{subsec:recovery}

\begin{figure}[t]
\centering
\small
\begin{lstlisting}
struct smallTy {
  int a, b, c, d, e;
};
struct largeTy {
  struct smallTy v;
  char pad[8];
};
int foo(struct largeTy *n) {
   return n->v.e;
}

Without recovery:
1. lea 0x10(%rdi), %rdi ;compute &n->v.e
2. mov $0x1FFFFFFFFFFFF, %r10
3. and %r10, %rdi  ;reset top 15-bits
4. mov (%rdi), %eax ;eax = n->v.e
5. ret

With recovery:
6. lea 0x10(%rdi), %rdi ;compute &n->v.e
7. mov $0x1FFFFFFFFFFFF, %r10
8. mov %rdi, %r11   ;saving the tag
9. and %r10, %rdi   ;reset top 15-bits
10.mov (%rdi), %eax ;eax = n->v.e
11.ret

Stub:
mov (%rdi),%eax  ;execute excepting instr
pop rdi          ;restore base register
ret
\end{lstlisting}

\caption{\toolname generates code labeled as ``With recovery'' (line:6-11) to recover from the size-invariant errors. In this case, \toolname does not add bounds-check before the memory access in 
{\tt foo}.
Instead, \toolname ensures that the pointer tag is live (line:8) across the memory access (line:10) to enable the emulation of the bounds-check in the signal handler, which is called if the size-invariant property is violated at runtime.}
\label{fig:recovery}
\end{figure}
\piyus{A legal memory access can cause an access violation if the size-invariant property is violated at runtime. We discuss our technique to recover from these errors using the example in Figure~\ref{fig:recovery}.} 

\piyus{Let us say the caller of {\tt foo} passes a single object of type {\tt struct smallTy}, and consequently, the argument {\tt n} gets invalidated because the size invariant property does not hold at the call site. However, access to {\tt n->v.e} in {\tt foo} is legal because it's within the bounds of the object (\texttt{pad} is outside the bounds).} 

\piyus{At function entry, {\tt \%rdi} contains the argument {\tt n}. At line-1, the address of {\tt n->v.e} is computed.} At line-3, \toolname resets the offset field in the pointer tag. At line-4, the actual dereference happens. Because the size invariant property does not hold, the hardware throws an exception at line-4. At this point, the signal handler in the userspace is called. To recover from fault, we perform a bound check in the signal handler 
for which we need to compute the base address. The base address can be computed using the fault address, tag bits, and the offset from the static-base. However, as we can see, the tag information is lost at this point because the {\tt \%rdi} register that was originally holding the tag has been overwritten at line-3. To obtain the tag bits, we have modified the compiler to ensure that the value of the potential fault address with the tag remains live during the access violation. In the modified assembly, at line-8, the compiler saves the content of {\tt \%rdi} in the {\tt \%r11} register, which is live during the memory access.

In addition, 
 \toolname generates metadata that is used by the signal handler to emulate the bounds check. The metadata includes the base register and displacement of the potential excepting instruction ({\tt \%rdi} and 0), the register that contains the tag ({\tt \%r11}), the offset from the static-base (0x10), the size of the memory access (4), and the length of the excepting instruction. Using this information, \toolname performs the bounds check in the signal handler. If the bounds check succeeds, \toolname generates a stub corresponding to the excepting instruction. The first instruction in the stub is the excepting instruction. The stubs are cached and reused for future faults to the same instruction pointer. Before calling the stub, the signal handler saves the address of the next instruction (address of line-11) and the contents of the base register ({\tt \%rdi}) on the stack (i.e., the stack pointer before the exception). It then sets the instruction pointer to the starting address of the stub and resets the invalid bit in the base register ({\tt \%rdi}) before returning from the signal handler. After returning from the signal handler, the stub code is executed that executes the excepting instruction and restores the value of the base register ({\tt \%rdi}) before returning to the original code (line-11).

This approach can only help us recover from those accesses 
in which the offset field in the tag is less than MAX\_OFFSET. We cannot retrieve the base address for large objects because the invalid bit is used for both size-invariant violation and an out-of-bounds address. The additional overheads for these changes are in the range 0-5\% for the SPEC benchmarks.

\subsection{Library Calls}
We assume that system libraries are safe. Since library code cannot interpret our tagged pointers, we add wrappers around library calls to mediate between an instrumented application binary and unmodified system libraries, as shown in Figure~\ref{fig:architecture}. We trust most of the library functions to use pointer arguments safely. For some library routines, we 
insert bounds check to ensure spatial safety.

For many library calls, \toolname 
simply resets the tags in the pointer arguments before calling the target function. 
However, this is not always sufficient. For example, library functions may return an interior pointer, perform a callback to the application routine compiled using \tname, and return their internal objects. In addition, the internal fields of an argument may contain tagged pointers. \toolname uses a custom implementation to handle these cases correctly.

\subsection{Object Initialization and Memory Accesses}
If an object is not initialized properly, the application may access any arbitrary memory location. 
To prevent such cases, we initialize the pointer fields in all allocations (including stack and global variables) with {\tt NULL}. Furthermore, if a global variable is initialized with an interior pointer, we also update the corresponding tag in the initialization.

If memory access is guarded by a bounds check, we reset the pointer tag before the memory access; otherwise, we only reset the offset field to catch the invalid accesses using pointers that do not satisfy the size-invariant (Section~\ref{sec:size_inv}).

For indirect calls with memory operands, we reset the pointer tag. 
If the address of a function leaves the static scope, we make it invalid. Marking the function addresses invalid disallows read/write on these addresses; however, the execution of invalid addresses is allowed using an indirect call. As a result, the application can execute any arbitrary virtual address using an indirect call. The existing mechanism for protecting indirect calls \cite{abadi2009cfi,zeng2011cfi} can be used alongside our scheme to enforce control flow integrity.
\section{Implementation}
\label{sec:implementation}
We implemented \tname \cite{safetypaper, paper1} as a compiler pass in the {\tt LLVM\--10.0.0} compiler. We extended the {\tt JEMALLOC\--5.2.1} allocator to allocate large objects from our segment-based allocator as discussed in Section~\ref{sec:design}. We discuss below our implementation and
optimizations to reduce the CPU overheads.

\paragraph{{\bf Instrumentation}} To insert the bounds check, we need to find the base first. If multiple memory accesses share the same static base, we try to insert a single call to our base finding routine that gets invoked at runtime. We implemented the base finding routine in the assembly. A function call may create register pressure, so we use a different calling convention for these calls that only uses the first argument (\%rdi) and the return value (\%rax) as caller-saved registers. In the fast path of our implementation,
only \%rdi and \%rax are used. In the slow path, we save/restore other registers that are used. Once we calculate the base, the bounds check logic, as shown in Section~\ref{sec:base_comp}, is directly instrumented in the LLVM IR.

\paragraph{{\bf Detecting illegal accesses}} Notice that in addition to bounds checks, we also rely on the invalid bit in the tag area to detect illegal accesses. Before accessing memory, we reset the top 15 bits excluding the invalid bit. If the invalid bit is set, the hardware generates the {\tt SIGSEGV} signal because the address is non-canonical. We register a signal handler for {\tt SIGSEGV} using the {\tt sigaction} system call. If an exception is generated due to illegal memory access, the operating system transfers the control to the signal handler in the user space. In the signal handler, we also implement our recovery mechanism for size-invariant errors.

\paragraph{{\bf Finding base}} 
The {\tt get-base-allocator} routine (Section~\ref{sec:design}), takes an internal address of an object and returns the base address of the object.
For heap objects, the base address is computed using the allocator's internal data structures.
For large-heap objects, the base address is computed using the alignment property of the segments. 

To support the base finding for stack variables, we register stack objects with the allocator when they are created and deregister them when they are destroyed. Note that this is required only for the stack variables that escape the static scope or are typecasted to an integer.
For global and static variables, at load time, the allocator walks global objects in different sections of the executable as specified in the executable format and stores the bases in sorted order to find the base using the binary search during the program execution.

\ignore{{\bf Variable-length arguments:} Our current implementation does not ensure safety for variable-length arguments. In our implementation, we have assumed that the variable number of arguments is always 16. To correctly handle this case, we require the caller to pass the number of arguments for every function call. For this purpose, we can use a spare caller-saved register, which is not used to pass the argument.}

\paragraph{{\bf Memory-mapped files and shared memory}} We do not support memory-mapped files and some shared-memory APIs. However, we support {\tt ANONYMOUS mmap} by allocating one extra page for storing the object-header. Notice that {\tt mmap} always returns a page-aligned address. 

\paragraph{{\bf Loop optimization}} If i) a pointer is always accessed inside a loop, ii) the pointer address only depends on the induction variable and the values outside the loop, iii) the lower bound, upper bound, and the step count of the induction variable are known, iv) the loop executes at least once, and v) the loop condition is the only way to exit from the loop, then we move the bounds check outside the loop.
The example in Figure~\ref{fig:loop_bounds} demonstrates our optimization.

\begin{figure}
\small
\begin{lstlisting}[language=C]
(*\bfseries Before optimization:*)
if (j > 0) {
  for (i = 0; i < j; i++) {
    bounds_check(arr_base, &arr[i+k], 
                 &arr[i+k+1], arr_limit);
    arr[i+k] = m;
  }
}
(*\bfseries After optimization:*)
if (j > 0) {
  bounds_check(arr_base, &arr[k], 
               &arr[k+j], arr_limit);
  assert(&arr[k+j] > &arr[k]);
  for (i = 0; i < j; i++)
    arr[i+k] = m;
}
\end{lstlisting}
\caption{{\tt arr} is accessed inside the loop. Since the lower and upper bounds for the array accesses inside the loop are {\tt arr[k]} and {\tt arr[k+j]}, the bounds check is moved outside the loop.}
\label{fig:loop_bounds}
\end{figure}

\ignore{
The example in Figure~\ref{fig:loop_bounds} demonstrates our optimization.
In this program, {\tt arr[i+k]} is always accessed inside the loop, i is the induction variable, and k and j are defined outside the loop. The lower bound, upper bound, and the step count of {\tt i} are zero, {\tt j}, and one. The loop executes at least once because {\tt j} is greater than zero.
In this case, we can statically compute the lower bound and upper bound of all possible accesses within the loop, i.e., {\tt arr[k]} and {\tt arr[k+j]}.
Thus, we can remove the check inside the loop and place one check outside the loop to check that memory accesses from {\tt arr[k]} to {\tt arr[k+j]} are safe.
To prevent the underflow and overflow of lower and upper bounds, we add an assertion that the upper bound is greater than the lower bound.
}
\paragraph{{\bf Updating pointer tag}} 
The tag update for escaping pointers (Section~\ref{sec:tag_ptr}) requires a bounds-check involving memory access. 
However, the escaping pointers could be invalid (or uninitialized) 
such as the ones pointing to a freed object, and may trigger a violation, if accessed. Even though \toolname handles these unlikely situations, we omit the implementation details since we never encountered this case in any of our benchmarks.
To recognize invalid pointers, \toolname initializes all pointers fields with NULL during allocation and marks all pointers initialized with a constant integer or a function address as invalid.
\section{Evaluation}
\label{sec:evaluation}
\subsection{Experimental Setup and Benchmarks}
We ran our experiments on a machine running Ubuntu-20.04.2 equipped with an 8-core 
3.6 GHz Intel i9-9900k processor, 32GB RAM, 1-Gb Ethernet controller, and 512GB SSD drive for persistent storage. We disabled hyper-threading during our experiments.
We measured CPU overheads using the SPEC CPU2017 \cite{spec2017} benchmarks. We used the reference input size for SPEC. For multicore performance, we used Phoenix-2.0 \cite{yoo2009phoenix} and the Apache-2.4.46 webserver. For Apache, we also instrumented {\tt apr-1.7.0} and {\tt apr-\-util-\-1.6.1} for spatial safety checks.
We configured Phoenix and Apache not to use memory-mapped files. In addition, we configured Apache to use \textit{anonymous MMAP} for shared memory instead of \textit{System V shared memory APIs}.
Phoenix \cite{yoo2009phoenix} reports that for the {\tt kmeans}, {\tt pca}, and {\tt histogram} benchmarks, the {\tt pthread} version is more scalable than the {\tt map-reduce} version. We ran the {\tt pthread} version for these benchmarks and the {\tt map-reduce} version for the rest. We used the large input set in our evaluations.
For {\tt matrix-multiply}, {\tt pca}, and {\tt kmeans}, we used input sizes of 2000x2000, 3000x3000, and 200000 respectively to make them run for at least a second. These benchmarks have a very short execution time even for the large input set.
For the security evaluation, we ran the BugBench \cite{lu2005bugbench} benchmark suite. 

To measure the execution time, we took the median of five runs for every benchmark. To report the memory overhead, we used the ``Maximum resident set size'' reported by ``/usr/bin/time -v'' command. We used the geometric mean to compute the average overhead. For the server experiment, we ran the client on a different machine (with a 1-Gbps network card) and directly connected both the machines. For scalability experiments, we disabled CPU cores using the CPU hotplug feature in the Linux kernel. For native results, we used the unmodified version of the {\tt LLVM-10.0.0} compiler and the {\tt JEMALLOC-5.2.1} allocator that we have used for our implementation.
\piyus{In our configuration for the native run, we used ``-fsanitize=address'' flag to generate the results for AddressSanitizer \cite{addresssanitizer}. We disabled the memory leak detection feature of AddressSantizer because we could not run some benchmarks due to memory leaks.}   
We compiled all our benchmarks with the O3 optimization level. The {\tt GeoMean} label in our graphs represents the geometric mean average.

\subsection{Performance}
\begin{figure*}[t!]
    \centering
    \begin{subfigure}{.57\textwidth}
    \includegraphics[width=\textwidth, height=4.5cm]{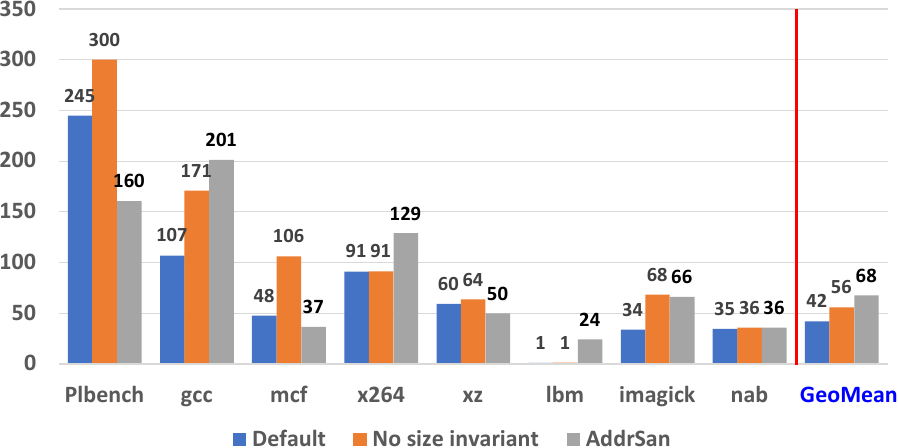}
    \caption{\% runtime overheads w.r.t.~native execution for \toolname with size-invariant optimization, \toolname without size-invariant optimization, and AddressSanitizer.}
    \label{fig:spec_cpu}
    \end{subfigure} \hfill
    \begin{subfigure}{.4\textwidth}
    \includegraphics[width=\textwidth,height=4.5cm]{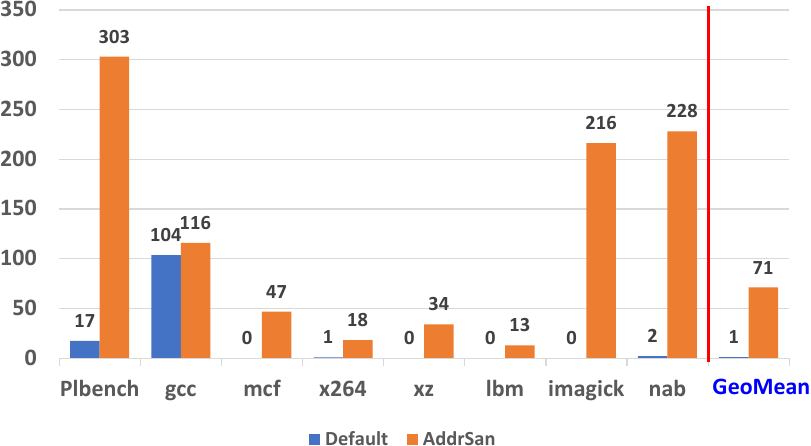}
     \caption{\% memory overheads w.r.t.~native execution for \toolname and AddressSanitizer.}
    \label{fig:spec_mem}
    \end{subfigure}
    \caption{\piyus{Runtime and memory overhead of \toolname and AddressSanitizer for the SPEC benchmarks.}}
\end{figure*}

Figure~\ref{fig:spec_cpu} shows the runtime overheads for SPEC benchmarks with and without size-invariant optimization, and the runtime overheads of AddressSanitizer. With all optimizations, the overheads are in the range of 1-245\%. The geometric mean 
is 42.1\%, as shown in the last column. {\tt Perlbench} (denoted as 
{\tt Plbench}) has the worst overhead of 245\%, whereas {\tt lbm} shows merely 1.2\% overhead. SGXBounds reported 41\% overheads inside the SGX enclaves \cite{SGX, mckeen2013innovativesgx} and 55\% overheads for outside the enclaves for SPEC CPU2006. Their average overhead also includes C++ benchmarks; therefore, direct comparison is not possible. Outside enclave, SGXBounds overheads for {\tt lbm} and {\tt mcf} are around -50\% (better than native) and 30\% compared to our overheads of 1.2\% and 47.9\% for these benchmarks. Inside enclave, SGXBounds reported around 5\% overhead for {\tt lbm} and 1\% overhead for {\tt mcf}. Interestingly, SGXBounds reported that {\tt lbm} also performs better than the native version for the AddressSanitizer implementation in the LLVM compiler. They attributed the change in memory layout to this speedup. {\tt Perlbench} and {\tt gcc} are the two worst performing benchmarks in our experiments. 
They were not evaluated by SGXBounds because they require custom modifications in the source code. We also require custom changes for these benchmarks, as described in Section~\ref{sec:usability}. The overheads of {\tt gcc}, {\tt mcf}, and {\tt imagick} are 170.9\%, 106\%, and 68.3\% without size-invariant optimization compared to 107.2\%, 47.9\%, and 34.1\% overheads with the size-invariant
optimization 
indicating its usefulness. 

\piyus{AddressSanitizer performs better than SGXBounds in an unconstrained environment \cite{sgxbounds}.
Therefore, we also compared the performance of \toolname with AddressSanitizer. AddressSanitizer could run all SPEC CPU 2017 benchmarks with the geometric mean overhead of 68\%. The overheads were in the range of 24\%-201\%. It 
performed better than \toolname for {\tt Perlbench}, {\tt mcf}, {\tt xz} benchmarks. Notice that AddressSanitizer had high overheads for {\tt Perlbench} (160\%, lower than \toolname 245\% overhead) and {\tt gcc} (201\%, higher than \toolname 107\% overhead) benchmarks. SGXBounds could not run these benchmarks. Unlike SGX\-Bounds, in our experiments {\tt lbm} did not perform better than native using AddressSanitizer.}

Figure~\ref{fig:spec_mem} shows the memory overhead of \toolname and AddressSanitizer for the SPEC benchmarks. Our memory overhead for SPEC is 1.16\% (Figure~\ref{fig:spec_mem}), which is slightly higher than the 0.4\% overhead reported by SGXBounds. {\tt gcc} and {\tt perlbench} are the worst-performing benchmarks with overheads of 104\% and 17\%, respectively. For {\tt gcc}, our memory overhead is mainly due to the source code modifications related to the size-invariant (discussed in Section~\ref{sec:usability}). To confirm this, we ran the native run with our custom allocator. 
The memory overhead in this case was 2\%. We performed a similar experiment for {\tt perlbench} and observed 16\% overhead. This confirms that source code refactoring is not the reason for the memory overhead in {\tt perlbench}. To validate that the overhead is not due to our segment-based allocation, we modified the original allocator to allocate eight additional bytes for every allocation. With the modified allocator, the overhead was the same as with our custom allocator. This indicates that the overhead is primarily due to the small objects for which the overhead of object headers is high. \piyus{On the other hand, the memory overheads of AddressSanitizer were in the range of 13\%-303\%, with the geometric mean overhead of 71.2\%. This is expected because AddressSanitizer uses shadow memory and puts freed objects into quarantine.} 

To summarize, it is difficult for us to directly compare with SGXBounds because they could not run {\tt Perlbench} and {\tt gcc}, which shows substantial overhead with our tool. \piyus{On average, \toolname performs better than AddressSanitizer, which performs better than SGXBounds in an unconstrained environment.}

\begin{figure*}[t!]
    \centering
    \includegraphics[width=\textwidth,height=4.5cm]{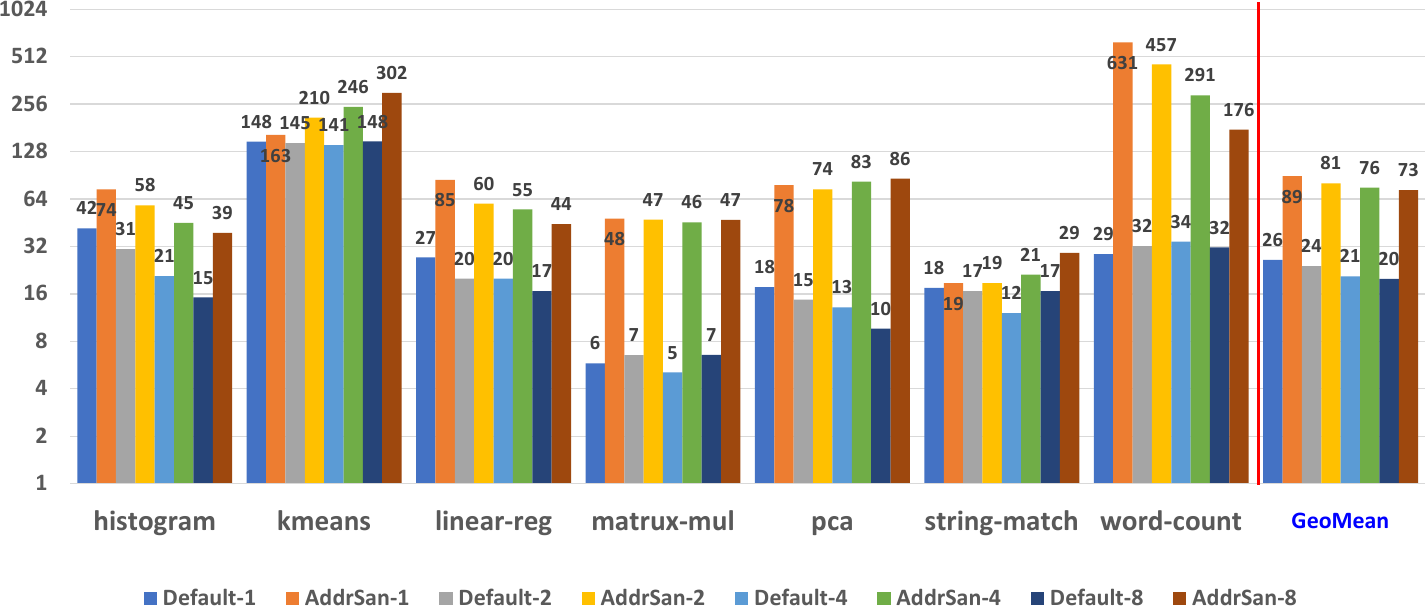}
    \caption{\piyus{\% runtime overheads w.r.t.~native execution of \toolname and AddressSanitizer for Phoenix benchmarks running on 1,2,4, and 8 CPUs. Default-n corresponds to \toolname with n CPUs, AddrSan-n corresponds to AddressSanitizer with n CPUs.}}
    \label{fig:phoenix_cpu}
\end{figure*}

\begin{figure*}[t!]
    \centering
    \includegraphics[width=\textwidth,height=4.5cm]{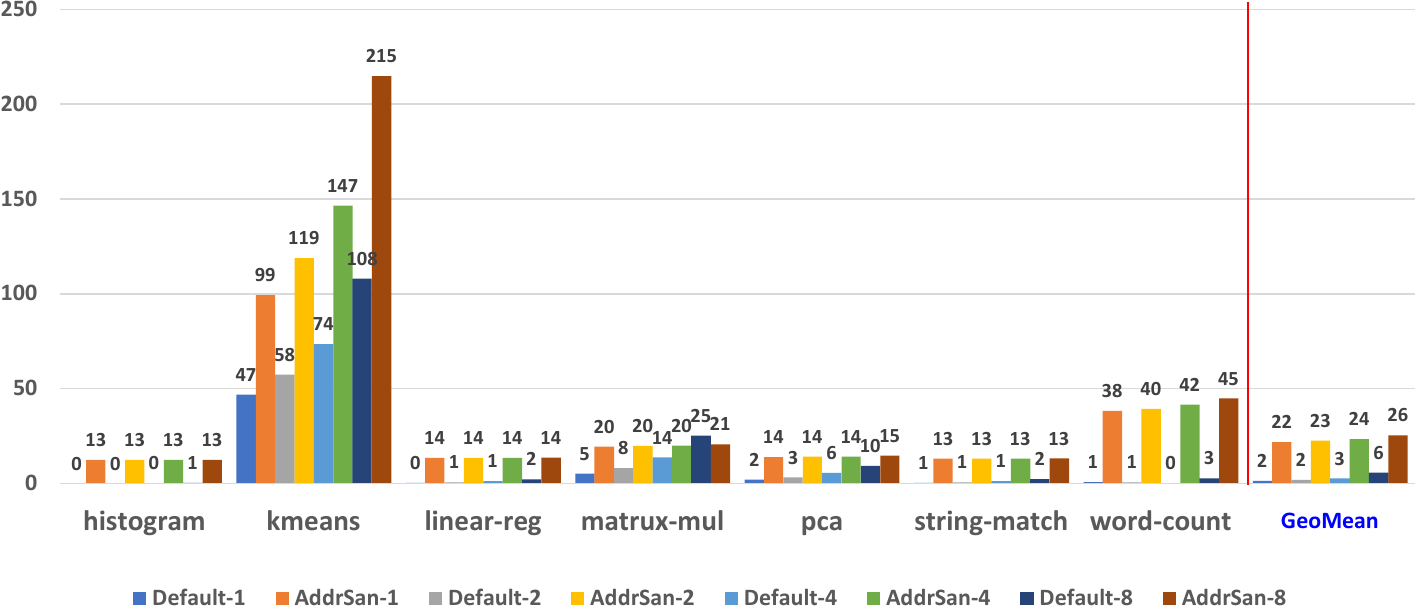}
   \caption{\piyus{\% memory overheads w.r.t.~native execution of \toolname and AddressSanitizer for Phoenix benchmarks running on 1,2,4, and 8 CPUs. Default-n corresponds to \toolname with n CPUs, AddrSan-n corresponds to AddressSanitizer with n CPUs.}}
    \label{fig:phoenix_mem}
\end{figure*}

\paragraph{\bf Scalability}
To test the scalability of our approach, we ran the Phoenix benchmark suite with 1, 2, 4, and 8 CPUs. Figure~\ref{fig:phoenix_cpu} shows the execution time overhead of \toolname and AddressSanitizer w.r.t. 
the native execution. Phoenix's average CPU and memory overheads using \toolname are 26.3\% and 1.6\% on a single core and 19.9\% and 5.9\% on eight cores respectively. As expected, our performance does not degrade significantly as the number of cores increases. However, we observed a sharp decrease in overheads with an increasing number of CPUs in the {\tt histogram} and {\tt linear-regression} benchmarks. This is because both of these benchmarks are not fully utilizing the CPUs on multiple cores, thus leaving scope for \toolname to steal some CPU cycles. The CPU utilization for {\tt histogram} for the native run on 1, 2, 4, and 8 cores is 99\%, 139\%, 177\%, and 205\%, compared to 99\%, 151\%, 205\%, and 251\% CPU utilization for \tname. A similar pattern is observed for the {\tt linear\--regression} benchmark, 
where the additional CPU overheads after disabling the size-invariant optimization were within the range of 10\% except for the {\tt kmeans} for which 
it is around 25\%.

For the Phoenix benchmark suite, SGXBounds performs better than \tname. For {\tt kmeans}, SGXBounds reported around 60\% overhead compared to 148\% overhead in our approach. For the remaining benchmarks, the CPU overheads in SGXBounds were less than 10\%. Notice that, unlike \toolname, SGXBounds does not need to update the tags on every pointer arithmetic. We discussed several optimizations in our design to address this issue. Still, for benchmarks in which the interior pointers are stored or passed to a function frequently, SGXBounds may perform better. However, unlike SGXBounds, our solution works even if the total memory consumption exceeds 4GB.  
\piyus{\toolname outperformed AddressSantizer for all benchmarks. The CPU overheads of AddressSanitizer were in the range of 18\%-631\%, with the geometric mean overheads of 89.4\% and 72.9\% on single and eight cores, respectively.}

We observed large variations in the memory overheads for the {\tt kmeans} and {\tt matrix multiply} benchmarks (Figure~\ref{fig:phoenix_mem}). The overheads vary between 47-108\% for {\tt kmeans} and 5-25\% for {\tt matrix\--multiply}. The memory consumption of these benchmarks is very small: 10MB for {\tt kmeans} and 53MB for {\tt matrix-multiply}. We believe that the page table pages corresponding to our custom heap segments are adding a few extra MBs, which is prominent due to the small memory footprint. To validate our hypothesis, we ran these benchmarks with relatively large inputs, and the resulting overheads of {\tt kmeans} and {\tt matrix-multiply} were in the ranges 57-62\% and 2-6\%. To further validate that the high overheads in {\tt kmeans} are not due to our segment-based allocation, we ran the native version with a modified allocator that allocates eight extra bytes for each allocation. In this case, we found that the memory overheads for {\tt kmeans} w.r.t. 
the native execution were in the range of 1-4\%. This means that the memory overheads in {\tt kmeans} are mainly due to the large number of small 
live objects. \piyus{AddressSanitizer also showed significant variations in memory overheads (99\%-215\%) for different numbers of CPUs for {\tt kmeans}. As expected, the geometric mean memory overhead of AddressSanitizer is higher than \tname.}
 
\begin{figure}
    \centering
    \includegraphics[width=0.45\textwidth,height=4cm]{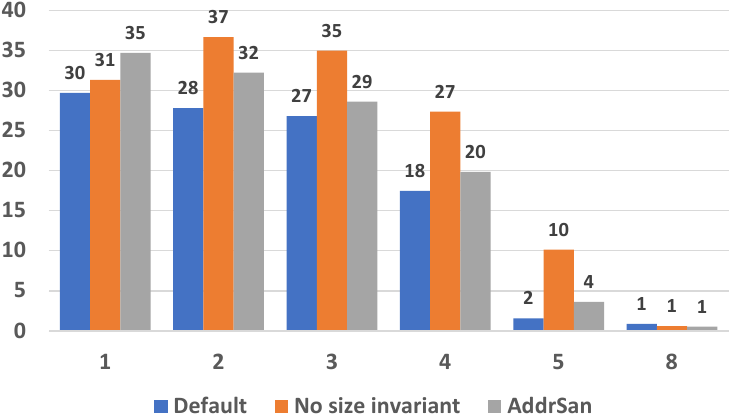}
    \caption{\piyus{\% reduction in the number of requests per seconds w.r.t.~native execution (\toolname w/ size-invariant optimization, \toolname w/o size-invariant optimization, and AddressSanitizer) for the Apache webserver running on 1,2,3,4,5, and 8 CPUs.}}
    \label{fig:apache}
\end{figure}

To further validate the usability of our tool for real applications, we ran the Apache webserver. Using a 1Gbps network card, we could not saturate all the cores even with concurrent requests. In the native run, the network card could only saturate three cores, so we ran our experiments with increasing number of cores. We ran the {\tt ab} tool on the client machine and enabled the {\tt KeepAlive} feature in the requests. To find the right metric for the concurrency level, we tried different parameters until we observed either a reduction or no significant change in the throughput. During these experiments, we ran our instrumented server and used 
its default pages. We got different concurrency levels for a different number of cores. 

Figure~\ref{fig:apache} plots the result for 1,2,3,4,5, and 8 cores. The first and second bars correspond to overheads with and without the size-invariant optimization. \piyus{The third bar shows the overheads of AddressSantizer. We observed 29.7\% overhead with the size-invariant optimization, 36.6\% overhead without the size-invariant optimization, and 34.7\% with AddressSanitizer when the CPUs were fully saturated (i.e., with less than four cores).} Our numbers 
started improving when the cores were partially saturated in the native run. With eight cores, we observed only 0.9\% overhead. The relative standard deviations 
were in the range of 0.25-1.47\% across all runs.

\subsection{Security}
\begin{table}
\small
    \centering
    \caption{BugBench benchmarks, which have known spatial safety bugs, SPEC CPU2017 and Phoenix-2.0 benchmarks, and known vulnerabilities in Nginx and Memcached applications for which \toolname could detect spatial safety violations.  The second column contains the filename:line-number pairs, at which the spatial safety violations were detected.}
    \begin{tabular}{|p{0.12\textwidth}|p{0.30\textwidth}|}
        \hline
         Benchmark & Access violation points \\
         \hline
         bc & bc.c:1425; util.c:270,577; storage.c:177 \\
         gzip & gzip.c:828 \\
         man & man.c:977,983,155; manfile.c:243 \\
         ncompress & compress42.c:896 \\
         ploymorph & polymorph.c:120,44,277,194,198,200, 231 \\
         \hline
         gcc & reload1.c:1868  \\
         x264 & biaridecod.c:297 \\
         \hline
         string & string\_match.c:158 \\
         \hline
         CVE-2013-2028 & ngx\_recv.c:136 \\
         \hline
         CVE-2011-4971 & memcached.c:3534 \\
         \hline
    \end{tabular}
    \label{tab:bugs}
\end{table}
To test the effectiveness of \tname, we ran the BugBench \cite{lu2005bugbench} benchmark suite, which contains a set of buggy applications some of which have spatial safety bugs. Table~\ref{tab:bugs} shows all the program points at which \toolname detected out-of-bounds accesses for the BugBench, SPEC, and Phoenix benchmark suites. We found all the bugs reported in the BugBench code repository.
In addition, \toolname also detected spatial safety violations in {\tt gcc} and {\tt x264} benchmarks from the SPEC CPU2017 benchmark suite. 

In {\tt gcc}, the global variable {\tt hard\_regno\_nregs} is accessed using a negative index. The check for the negative index is conducted after the variable access. Importantly, the AddressSanitizer implementation in LLVM could not detect this bug in {\tt gcc}. SPEC CPU2017 contains an old version of {\tt gcc} compiler. This bug is not present in the current {\tt gcc} compiler.

In {\tt x264}, global variables {\tt INIT\_FLD\_MAP\_I} and {\tt INIT\_FLD\_LAST\_I} are accessed at an index that is outside the bounds of these objects. These variables are passed at lines-90,91 in {\tt context\_ini.c}. 
In the {\tt string\-\_match} benchmark from the Phoenix, {\tt fdata\_keys}, which is allocated for size {\tt finfo\_keys.st\-\_size} at line-259 in {\tt string\-\_match.c} is accessed in the loop. This loop has an incorrect bound check in the loop condition that allows the program to access an additional byte past the original allocation size.
Our post-evaluation inspection revealed that the bug reported for {\tt perlbench} \cite{AddressSanitizerBugs} in SPEC CPU2006 has already been fixed in SPEC CPU2017. Therefore, \toolname did not report it.

\toolname also detected known spatial safety violations in {\tt nginx\--1.4.0} (CVE-2013-2028)\cite{CVE1} and {\tt memcached-1.4.4} (CVE-2011\\--4971)\cite{CVE2}. 
Nginx was tricked into receiving a message of arbitrary length, which can be controlled by the user in a stack-allocated array. The spatial safety checks in our library wrapper 
for {\tt recv} caught this bug. In Memcached, a negative value is passed as a size parameter to the {\tt memmove} routine 
triggering violation. 

To summarize, \toolname could detect all the bugs reported by existing tools and found a new bug in the {\tt gcc} benchmark from SPEC CPU2017.


\subsection{Usability}
\label{sec:usability}
\begin{table}
    \small
    \caption{Source code refactoring: Type a) changes related to size-invariant, b) automatic pointer comparison to int comparison generated by the frontend, and c) other changes. 
}
    \begin{tabular}{|p{0.07\textwidth}|p{0.02\textwidth}|p{0.27\textwidth}|p{0.04\textwidth}|}
        \hline
         Benchmark & Type & Source code modification & KLOC \\
         \hline
         \multirow{4}{*}{Perlbench} & a & hv.h:48; pad.c:2808; MD5.c:184; op.c:8401 &  \\\cline{2-3}
         & b & regexp.h:474, regcomp.c:13764 & 291 \\\cline{2-3}
         & c & pp\_pack.c:3038; av.c:159; perly.c:408; pp\_hot.c:3175; regcomp.c:16274; & \\
         \hline
         \multirow{3}{*}{gcc} & a & tree-ssa-operands.c:130,133; tree-ssa-sccvn.c:1542,1580,1610; tree.c: 2102,863,865,958,1467,1584,3604, 9411; sbitmap.c:82; gimple.c:148; sparseset.c:38; rtl.c:199,341; reload1.c:915; cpp\-\_symtab.c:173 & \\\cline{2-3}
         & b & obstack.h:526,538 & 971 \\\cline{2-3}
         & c & c-common.c:5296; dominance.c:1339; ggc-page.c:571; pointer\-\_set.c:67 & \\
         \hline
         Apache & a & event.c:1501 & 301 \\
         \hline
         \multirow{2}{*}{Phoenix} & b & linear\-\_regression.c:256 & 7 \\\cline{2-3}
         & c & atomic.h:81 & \\
         \hline
    \end{tabular}
    \label{tab:refactor}
\end{table}

For most benchmarks, we did not need to refactor source code. Table~\ref{tab:refactor} provides a summary of our changes. At a broad level, we
categorized the changes as follows: 
a) 
Related to the size-invariant, b) 
Related to a pointer comparison converted to an integer comparison by the frontend, and c) Other. 

Most changes were related to the size-invariant, and these were the easiest to fix. We found that for most of these cases, \toolname threw an exception at the allocation point itself. 

In some cases, the frontend generated an integer comparison instead of a pointer comparison. 
These conversions were typically done for ``!'' style comparison. We refactored the code 
so
that the frontend generated a pointer comparison. In the future, we plan to extend the frontend to avoid the need for these changes. 

In {\tt gcc}, pointers are used as integers 
in comparisons, array indexes, and hash table keys.
In all these cases, we changed the source code to reset the pointers' tags.

\ignore{To summarize, most benchmarks did not require any refactoring. Even for large applications e.g. Apache, we needed refactoring at only one place. This indicates that our technique 
is feasible.
We also ran {\tt gcc}, {\tt perlbench}, and {\tt apache} without the size-invariant modifications. \toolname could successfully run {\tt perlbench} and {\tt apache} without any additional overheads. This is because the parts of code that require size-invariant modification are not on the hot path. However, we could not run {\tt gcc} because it uses a custom allocator that uses system allocator in the backend. As a result, most of the objects are large for which \toolname cannot retrieve the base address during the size-invariant violations.}
To summarize, most benchmarks did not require any refactoring. Even for large applications e.g. Apache, we needed refactoring at only one place. This indicates that our approach is feasible. 
We tried to run {\tt gcc}, {\tt perlbench}, and {\tt apache} without the size-invariant modifications to test our size-invariant recovery mechanism. 
{\tt Perlbench} and {\tt apache} ran successfully without additional overheads because parts of code that require size-invariant modification are not on the hot path. However, {\tt gcc} crashed because it employs a custom allocator that uses system allocator in the backend. As a result, most of the objects are large for which \toolname cannot retrieve the base address during the size-invariant violations. 


\section{Limitations and future work}
\label{sec:limitations}
\toolname relies on a programmer to typecast an integer to a pointer if an integer with an inconsistent tag escapes the static scope and
is accessible later. 
In our experiments, we found that this practice is generally followed (see Section~\ref{sec:usability}). 
We also assume that the implicit integer-to-pointer typecasts are safe. In a rare case, if the size-invariant property is violated due to an implicit typecast some bugs may remain undetected.  

At a more general level, \toolname assumes that the developer's intent is 
benign. It also assumes a weaker form of type safety (discussed in the previous paragraph) and temporal safety. Existing works 
have similar limitations. In SGXBounds\cite{sgxbounds} approach, if the limit of the tagged pointer is modified using an integer, the bounds check may incorrectly succeed or fail at runtime. For BaggyBounds \cite{Baggy}, PAriCheck \cite{ParicheckAlloc}, and Low Fat Pointers \cite{LFPCC, LFPNDSS}, an out-of-bound pointer can be created and accessed using integer arithmetic. These works also require source code refactoring. The primary reason behind such limitations is that it is hard to statically track arithmetic operations on an escaped integer that is also a pointer.

Our current implementation does not ensure safety for variable-length arguments. To correctly handle this case, we require the caller to pass the number of arguments for every function call.

In the future, we will investigate whether our work can be extended to support temporal safety. However, existing techniques \cite{boehm1988garbage} for temporal safety can be used alongside our approach with minor modifications for tagged pointers. We also plan to extend \toolname to support an OS kernel.
\section{Related work}
\label{sec:related}
Jones and Kelly \cite{joneskelly} proposed the idea of object-bounds protection. However, it did not allow the creation of an out-of-bounds pointer. CRED \cite{CRED} improved 
on this work by 
supporting an in-bounds pointer derived from an out-of-bounds pointer. However, both these works suffered from CPU overheads due to the splay-tree-based implementation for bounds checking. Dhurjati and Adve \cite{CRED++} 
reduced CPU overheads by using per-pool splay-trees instead of a global splay-tree. 

Baggy Bounds \cite{Baggy}, PAriCheck \cite{ParicheckAlloc}, and Low Fat Pointers \cite{LFPCC, LFPNDSS} further reduce the CPU overheads by adding extra padding to objects that allow them to locate the base address without an expensive search. However, these works do not provide precise object-bounds protections because they allow the applications to access the padded area. These works have also used the pointer tagging approach. SGXBounds \cite{sgxbounds} provides precise object-bounds protection but restricts the application address space to 32-bit on a 64-bit platform. Delta Pointers \cite{delta} further reduces the CPU overheads of SGXBounds by only detecting overflows. \piyus{Delta Pointers can support 48-bit address space provided the maximum object size is restricted to 32 KB.} Both SGXBounds and Delta Pointers use pointer tagging, and they store the tag in the virtual address of the pointers, similar to us.
\piyus{CUP \cite{cup} uses a table to compute the bounds of an object at runtime. Because the table size can be huge, it limits the total number of objects to $2^{31}$ to restrict the table size and enable fast indexing. The table index is embedded in the address of an object.}

Another line of work provides spatial safety for pointer-bounds. These approaches can detect sub-object overflow at the cost of high CPU and memory overheads because they need to store and update bounds for every pointer.

CCured \cite{ccured,condit2003ccuredreal} statically categorized the pointers into SAFE, SEQ, and WILD. SAFE pointers are normal pointers and do not require any checks. SEQ and WILD pointers are fat-pointers that store the bounds information of pointers and objects and require runtime checks. Cyclone \cite{cyclone} uses fat-pointers and also provides programmers a variety of pointer qualifiers to control the runtime checks. SoftBound \cite{softbound} stores per-pointer metadata in a disjoint address space for better compatibility.
SafeC \cite{SafeC} and Xu et al. \cite{xu2004efficient} also track bounds for every pointer and can also detect temporal safety bugs in addition to spatial safety bugs. 

AddressSanitizer \cite{addresssanitizer}, Valgrind \cite{valgrind}, and Purify \cite{purify} can only detect sequential buffer overflows and underflows, but they also have much wider goals.
\section{Conclusion}
\label{sec:conclusion}
We presented \tname, a tool that provides precise object-bounds protection for C applications at low CPU and memory overheads without restricting the application address space on the x86\_64 platform. 
\toolname requires applications to obey a weak form of type-safety. Our evaluation revealed that for most applications, this property holds. The changes needed for applications that did not satisfy the property were minor. \toolname was able to detect spatial safety violations in widely used benchmarks. In particular, it detected a bug in {\tt gcc} that was not reported in any other works to the best of our knowledge. This evaluation demonstrates that our approach is effective and can scale to real applications.

\bibliographystyle{plain}
\bibliography{ms}

\end{document}